\documentstyle[prl,aps]{revtex}
 \twocolumn
 \draft
 
\begin{document}
\title{Squeezing Atomic Vibrations In An Optical Lattice. 
A New Mechanism Of Optical Cooling.}  
\author{ A. L.Burin$^{1,3}$, Joseph L.Birman$^{1}$, A.Bulatov$^{2}$,  and 
H.Rabitz$^{2}$ }
 \address{$^1$ Department of Physics,
 City College of the City University of New York, 
 138th Str. at Convent Ave., New York, N.Y. 10031; 
 \and \\  
 $^2$ Department of Chemistry, Princeton University, Princeton, NJ 08544, 
 \and \\
 $^3$ Current address: Department of Chemistry, Northwestern University, 
 Evanstone, IL, 60208
 }
\maketitle

%\ $^1$ Department of Physics, City College of the City University of New        York,  138th %Str. at Convent Ave., New York, N.Y. 10031

%\ $^2$ Department of Chemistry, Princeton University, Princeton, NJ 08544

%\ $^3$ Current address: Department of Chemistry, Northwestern University,        Evanstone, IL, %60208

\date{\today}
\maketitle
\begin{abstract}
We propose a new method to obtain a squeezed matter field of atomic
vibrations by use of an optical lattice, and the laser pulse technique of Garrett et al used for acoustic phonons [1]. 
We show that it is possible to reduce the variance of atomic momentum to a
value 
as low as the recoil momentum. Consequently, cooling to recoil energy 
can be achieved.  
\end{abstract}
\vspace{3mm}

It this letter we propose and analyze the possible squeezing of a matter
field comprising the atomic vibrations of an optical lattice (see e.g.
Ref.\cite{Muller}) which is created when atoms are spatially localized in
the potential minima of a standing laser field at ultra-low temperature.
As shown below this approach has major advantage compared to the phonon case
\cite{Merlin}. For a realistic system, we show that significant squeezing
can result, and that optical cooling can be achieved via this new method.
An analogous but different method was recently proposed in Ref.\cite{Ammann}
where it was assumed that the optical lattice is suddenly turned off and on.
In our case, the background optical potential is always present but there
exists an additional fast time-dependent component which leads to the
squeezing of the phase distribution of atoms in the optical lattice.

It is known that the squeezed state \cite{Review} can be achieved for a
Bose system with equally spaced oscillator levels, such as photons
\cite{Photon_Squeezing}, polaritons \cite{Polaritons}, and phonons
\cite{Polaritons}. 

In the recent experiment \cite{Merlin}, the squeezing of phonons in $KTaO_{3}$ 
has been achieved with a noise reduction of $10^{-4}$. This small value is due to several
factors limiting the squeezing that can be achieved by lattice phonons :
phonon dispersion which leads to the decay of the induced coherent
oscilations, residual anharmonicity (phonon-phonon interactions), finite
pulse length of the incident  laser pulse excitation. In the experiment
\cite{Merlin} some compensation of these negative factors occured by
tuning to a van Hove singularity in the two-phonon density of states. We
were motivated to consider an optical lattice because the characteristic
atomic vibration energy \cite{?} is some $10^{4}$ times less than for
atoms in a vibration (phonon) mode in a "real" lattice. Additionally the
phonons have essentially an Einsteinian (dispersionless) distribution.
The method discussed here is thus free from the above limitations which leads to larger squeezing. 

As a model of the optical lattice we consider a  gas of two level atoms
each of mass $M$ and resonant energy $E_{0}=\hbar\omega_{0}$, with transition dipole moment $d$ and where the radiative width of the excited state
is 
$\gamma_{R}$. The atoms of this gased are located in the minima of the
potential relief formed by the external laser standing wave which can be
characterized by the frequency $\Omega$ and the wave vector $k\approx
\omega_{0}/c$.  We assume that the detuning $$\Delta \omega = \omega-\omega_{0}$$ is
larger than the Rabi frequency $\Omega_{R}$ 
and the radiative width. The off-resonance conditions are required to
avoid decay of the coherent state due to the presence of the excited
state. For simplicity we consider a one dimensional model and the extension
to three dimensions is straightforward because of the separation of the
variables. 
The use of time-dependent off-resonant dipole potentials was shown to
be successful for the manipulation of the atomic center of mass motion in
optical
lattices \cite{Raizen}. 

Consider first the states of 
the atom neglecting the radiative width in comparison with the detuning
and the 
Rabi frequency. The energies of two eigen states can be found within the
standard 
$u-v$ transformation for two levels. All our work is made in 
the rotating frame. Then for zero Rabi frequency the ground level 
has energy $0$ and the excited level has complex energy 
$\Delta\omega + i\gamma_{R}$ which means that it has finite lifetime. 
For finite Rabi frequency $\Omega_{R}$ the hybridization of these levels
must be taken into account. 
Since the Rabi-frequency is spatially dependent in the field of the
standing wave we can represent 
the effective potential for the pseudoground state as \cite{Graham} 
\begin{equation}
U(x) \approx - \frac{\hbar\Omega_{R}^{2}}{2\Delta\omega}cos^{2}(kx) = 
- \frac{\hbar\Omega_{R}^{2}}{4\Delta\omega}(1+cos(2kx)), 
\label{eq:potential}
\end{equation}
where $k$ is the resonant wavevector. 
The last approximation implies an adiabatic condition, namely  
the vibrational motion within the optical lattice is much slower 
that the motion between two levels of a single atom. The   
frequency of the atomic vibrations near 
the potential minima in the effective potential (\ref{eq:potential}) 
is given by    
\begin{equation}
\omega_{h} = \sqrt{\frac{\Omega_{R}^{2}\hbar (2k)^{2}}{4M\Delta\omega}}, 
\label{eq:oscfreq}
\end{equation}
while the motion between the upper and lower levels is defined by
$\Delta\omega$.  
The energy 
\begin{equation}
E_{R} = \frac{\hbar^{2}k^{2}}{2M}
\label{eq:recoil}
\end{equation}
corresponds to the atom having recoil momentum $q = \hbar k$ and represents the
recoil energy $E_{R}$ which is normally much less than the radiative
width $\hbar\gamma_{R}$ and the detuning 
$\Delta\omega$ in the off-resonant case under consideration.

In analysing the optical lattice, we will assume the same type of
excitation as in \cite{Merlin}, namely a "delta" laser pulse
detuned from resonance of the two level atoms. In the work \cite{Merlin},
a rapid   
laser pulse, with duration $\tau \sim 70 fs$ less than the inverse phonon Debye
frequency was applied at $t=0$, perturbing the system by the potential: 
\begin{equation}
V(x) = -\alpha I x^{2},
\label{eq:interaction}
\end{equation} 
where $x$ is the displacement, I is the intensity of the pulse and
$\alpha$ is a proportionality factor which is unknown for a crystal
and will be found below in the 
optical lattice. Taking into account the small value of pulse duration
$\tau$ we can 
represent the time dependent perturbation by a  $\delta$ function 
\begin{equation}
V(x) = -\alpha I x^{2}\tau\delta(t),
\label{eq:interaction1}
\end{equation} 

Linear terms in $x$ are absent due to symmetry  \cite{Classic_Book}. The
system eigenfunction just after the perturbation ($t = 0^{+}$) is given
as 
\begin{eqnarray}
\Psi(x, t\rightarrow 0^+) = \Psi(x, t\rightarrow 0^-)\cdot e^{i\xi
M\omega_{h} x^{2}},
\nonumber\\
\xi = \frac{\alpha I \tau}{\omega_{h} M}, 
\label{eq:perturbation_of_phase}
\end{eqnarray} 
  
We have introduced the dimensionless parameter $\xi$ which characterizes
the strength of the squeezing. In order to examine the amount of
squeezing, we need to calculate the dispersions $p^{2}$ and $x^{2}$ in
momentum and configuration space respectively. Following \cite{Merlin} we
obtain 
\begin{eqnarray} 
<p^{2}>_{min} = <p^{2}>_{0-}\frac{1}{2\xi^{2} + 1 + 2\sqrt{\xi^{4} +
\xi^{2}}},
\nonumber\\
<p^{2}>_{max} = <p^{2}>_{0-}(2\xi^{2} + 1 + 2\sqrt{\xi^{4} + \xi^{2}}),  
\nonumber\\
<x^{2}>_{min, max} = <p^{2}>_{min, max}/(M^{2}\omega_{h}^{2}),
\label{eq:squeezing_effect}
\end{eqnarray} 
Where $<p^{2}>_{min}$ and $<p^{2}>_{max}$ correspond to the maximum 
and minimum values of the momentum dispersions due to the vibrations of 
the optical lattice, and $<p^{2}>_{0-}$ is the momentum dispersion of the
atoms in the lattice before the external pulse given by
Eq.(\ref{eq:interaction1}) was applied to the system. 
The product of the minimum and maximum variances of $p$ and $x$,
respectively, remains the same as before switching on the external pulse.
This conservation reflects the special properties of the harmonic oscillator
potential. 

For strong squeezing the result (\ref{eq:squeezing_effect}) can be
simplified as 
\begin{equation}
<p^{2}_{min}> \approx <p^{2}>_{0-}\cdot\frac{E_{i}}{E_{f}}
\label{eq:convsqueezing}
\end{equation}
where $E_{i,f}$ are the initial and final energies of the  oscillatior,
respectively. 

Our goal is to estimate the minimum variance of momentum which can be
achieved in the cooling scheme described above. This corresponds to the
maximum value of squeezing for the atoms 
forming the optical lattice. 
  
According to Eq.(\ref{eq:convsqueezing}) the minimum 
width of the wave function in momentum space can be represented as
$2ME_{i}^{2}/E_{f}$  . Let the depth of the well created by the standing
wave be $U$. The energy $U$ defines the maximum energy which can be
transmitted to the atom. The minimum initial energy 
$E_{i}$ is just the level spacing for the oscillator-type 
equidistant levels near the bottom of the well:   
$$
\hbar\omega_{h} \sim \sqrt{E_{R}U}.
$$
Taking into account all the above estimates we find that the minimum
possible width in momentum space which can be reached for the optical
lattice  in this method   is limited by 
the recoil momentum 
\begin{equation}
<(\Delta p)^{2}>_{infim} \approx (\hbar 2k)^{2}. 
\label{eq:absmin}
\end{equation}
This is the maximum possible squeezing which can be obtained without
spontaneous emission since $2k$ is the minimum momentum transfer for the
interaction with the laser field.  
Below we will study whether this strong squeezing can be achieved for 
real conditions. 

One should note that the anharmonicity might produce a further
constraint for the width of the squeezed state. Additionally it gives
rise to the slow decay of the amplitude of the variance oscillations.
However under the current experimental conditions \cite{Raizen} this
effect will change $<(\Delta p)^{2}>_{infim}$ by a factor of the order of
unity at least for the first vibration. A more detailed account of the
influence of anharmonicity will be published separately \cite{exp}.

The adiabatic treatment of the atomic 
motion in the potential of the laser field is   valid automatically since
the 
detuning $\Delta\omega$ is assumed to be much greater 
than both the recoil energy and the energy corresponding to the Rabi
frequency.  

In the off-resonant case $\Omega_{R} \ll \Delta\omega$ the decay rate of
the g-state can be written as 
(see e.g. Ref.\cite{Cohentannoudgi_book})
\begin{equation}
\gamma_{g} \approx \gamma_{R} \frac{\Omega_{R}^{2}}{\Delta\omega^{2}}.
\label{eq:lowdecay}
\end{equation}

Any decay event leads to a change of the phase of the atom wave-function
as well as 
the momentum of the 
atom. The loss of coherence might occur and squeezing will 
be hard to achieve. That is why we have to avoid   spontaneous emission.
Since   squeezing occurs during the period of the atom vibration, defined
by its inverse
frequency $(\omega_{h})^{-1}$, this frequency should be much larger 
then the decay rate (\ref{eq:lowdecay}) 
\begin{equation}
\gamma_{g} \ll \omega_{h}.
\label{eq:coher}
\end{equation}

We will consider the case of two fast waves counter-propagating, so:
\begin{eqnarray}
E(x, t) = E_{o}(cos(\omega_{*}(t-x/c))\varphi(\frac{t-x/c}{\tau})+
\nonumber\\
+ cos(\omega_{*}(t+x/c))\varphi(\frac{t+x/c}{\tau})),
\label{eq:extpulse}
\end{eqnarray}
where $\tau$ is the duration of the pulse,  
the function $\varphi(a)$ decreases rapidly for $a > 1$ and for detailed 
estimations we will use the Gaussian wave packet  
\begin{equation}
\varphi(a) = e^{-a^{2}/2}.
\label{gauss}
\end{equation} 
We use two waves for the pulse because in the case of one wave a linear
term appears in the phase of the excited atomic wave function. This term
does not influence  squeezing in a harmonic well, although the
anharmonicity might be more important in this 
case. All calculations can be easily justified for  
the more complicated picture of one fast pulse but there will be no 
difference for the maximum amount of squeezing. 

By "fast pulse" we mean that the atoms remain static during the pulse.
Thus the pulse duration $\tau$ must be much less than the period of
atomic vibrations in the well 
\begin{equation}
\omega_{h}\tau \ll 1.
\label{eq:fastpulse}
\end{equation}
To make the effect strong we need to have resonant conditions between the
external pulse and the stimulated oscilations of the atom dipole moment.
For this purpose the frequency of the pulse must be close to the
frequency of the standing wave field $\omega_{*}\approx \omega$. 

As in Ref.\cite{Merlin} the synchronization of disturbed vibrations of
different atoms should be provided. Consequently the distance passed by
the light during the characteristic period of the atomic vibrations
$2\pi\omega_{h}^{-1} \sim 10^{-9}s$ must be larger than the sample size.
The sample size should be less than ~1cm. We
assume this condition to be satisfied. Hence we can neglect x-dependent part
of the $\phi$-function argument in Eq.(\ref{eq:extpulse}).  

Consider the phase shift caused by the action of the external field on
the single atom. 
The effective potential acting on an atom can be represented as 
\begin{equation}
U_{tot} = U_{0}+\delta U = \frac{\hbar}{4\Delta\omega}(\Omega_{R}^{2} + 
\delta\Omega_{R}^{2}\tau\delta(t))cos^{2}(kx), 
\label{total_potential}
\end{equation}
The time dependence of the external pulse has been replaced with the
$\delta$ function and the perturbation of the Rabi frequency is defined
as 
\begin{equation}
\delta\Omega_{R}^{2} = d^2E_{0}^{2}\int_{-\infty}^{+\infty}\varphi(x)dx.
\label{rabi_perturbation}
\end{equation}

The phase shift caused by this external pulse is 
\begin{equation}
\Phi(x) \approx \int dt \delta U(x,t) =  -
\frac{I_{1}}{I_{0}}\frac{\Omega_{R}^{2}}{4\Delta\omega}\tau \cdot
2k^{2}x^{2}
\label{phase_shift}
\end{equation}
where $I_{0}$ is the intensity of the standing wave and $I_{1}$ is the
intensity of the pulse. The squeezing parameter $\xi$ (\ref{eq:perturbation_of_phase}) reads 
\begin{equation}
\xi = \frac{I_{1}}{I_{0}}\frac{\Omega_{R}^{2}}{8\Delta\omega}\tau \cdot
\frac{E_{R}}{\hbar \omega_{h}}. 
\label{eq:xi_optical}
\end{equation}

The validity of Eq.(\ref{eq:perturbation_of_phase}), i.e. that only a
change of phase occurs in $\Psi$, can be checked by estimating the
probability of the ground state $\rightarrow$ excited state transfer
under the pulse (\ref{eq:extpulse})given by the matrix
element 
\begin{equation}
M(x) \sim \frac{dE_{0}}{\hbar}(\eta((\omega_{*} - \omega_{0})\tau)) + 
\eta((\omega_{0} - \omega_{*})\tau)))
\label{eq:applic2}
\end{equation}
The requirement that the perturbation be sufficiently weak implies that this matrix
element must be much less than unity. For the case of a Gaussian
perturbation this condition can be rewritten as 
\begin{equation}
\frac{dE_{0}\tau}{\hbar}\sqrt{2\pi}exp(-\frac{(\omega_{0} -
\omega_{*})^{2}\tau^{2}}{2})\ll 1.
\label{eq:applic3}
\end{equation}
The exponentially small value of the excitation probability for the large
difference between 
$dE_{0}$ and $\Delta\omega$ is the general consequence of the
adiabatically slow change of the external field with respect to the
energy difference between levels. 

Thus we find that  the frequency of the external pulse perturbation must
be close to the frequency of the laser responsible for the formation of
the optical lattice and the duration of the short pulse should exceed the
inverse frequency of the transition between the ground and excited levels

\begin{equation}
\tau\Delta\omega \gg \hbar.
\label{eq:condadiab}
\end{equation}
For future consideration we will simply assume that the frequency of the
external pulse coincides with that of the standing wave field $\omega =
\omega_{*}$.

Using conditions  (\ref{eq:coher}),  (\ref{eq:condadiab}), 
we can examine whether the maximum squeezing, defined the by recoil momentum
of 
Eq.(\ref{eq:absmin}), can be attained, by estimating the largest pulse
duration $\tau$ and   pulse amplitude $E_{0}$.  

We take for the radiative width 
\begin{equation}
\gamma_{R} \sim 10^{8} s^{-1},
\label{eq:radwid}
\end{equation}
as is appropriate for alkali atoms \cite{RECENT}. The frequency
corresponding to the recoil energy is usually about $10^{2}$ times less
than the radiative width 
$$
\frac{E_{R}}{\hbar} \sim 10^{6}s^{-1}, \hspace{3mm} E_{R} \sim 10^{-9}eV.
$$

It is significant that the energy $E_{rad}=\hbar\gamma_{R}$ defines the
minimum  temperature   for the atoms   within the commonly adopted
Doppler cooling technique \cite{Cohentannoudgi_book}. Recall that to reach the minimum width in the momentum space we need to
have the initial energy of the order of the quantization energy $\omega_{h}
\sim \sqrt{UE_{R}}$ (see Eq.(\ref{eq:absmin}) and the preceding discussion).
Therefore the depth $U$ of the well in the standing wave required to get
the maximum squeezing must exceed $10^{-5}eV$.  This value of the well
depth   can be reached within existing experimental techniques (see
Ref.\cite{RECENT}). Note that the above set of parameters corresponds to
the limit of validity for small decay condition (\ref{eq:lowdecay}). 

The requirements for the external pulse are of most interest. To
reach maximum squeezing in accordance with Eqs. (\ref{eq:absmin}),
(\ref{eq:squeezing_effect}), (\ref{eq:perturbation_of_phase}) we need to
have 
\begin{equation}
\xi^{2} \sim \sqrt{\frac{U}{E_{R}}} \approx 10.
\label{eq:xireq}
\end{equation}
Using the definition of  $\xi$ Eq. (\ref{eq:xi_optical}) we get 
$$
\xi \sim \frac{I_{1}}{I_{0}}(\Omega_{h}\tau) \sim 10.
$$   
The latter condition implies that the amplitude of the field in the fast
pulse must be at least about 3-5 times larger than in the standing wave
field. This restriction does not appear to be crucial for the development of the
squeezing technique. 
Note that the "fast'' field remains adiabatically slow with respect to
the motion between interatomic levels and the condition
(\ref{eq:condadiab}) as well as its possible 
generalizations are satisfied. Then if these conditions are satisfied the
minimum momentum variance can be reached. For an alkali atom optical
lattice this will give squeezing of 10 compared to $10^{-4}$ for the
measured phonon case \cite{Merlin}. 

As regards cooling, recall that $T_{Dopp}\sim \hbar \gamma_{R}/k_{B}$
defines the minimum temperature whcih can be reached for the atoms with the
most common Doppler cooling technique 
\cite{Cohentannoudgi_book}. In the scenario proposed here the
laser is abruptly switched off just when the system is in the
state of maximum squeezing Eq.(\ref{eq:absmin}) (minimum variance) in
momentum space. As shown in Eq.(\ref{eq:absmin}) this is the recoil
momentum. The ratio of Doppler and recoil temperatures is about 100 in
alkali metals. Hence our method ensures sub-doppler cooling 100 times
down the Doppler limit $\sim 10^{-5}K$ during a relatively short time
$10^{-8}s$. Note also that the method permits multiple repetition (laser
on, laser off), and this cycling will be advantageous in reaching very
low temperatures. An approach to the cooling
problem in case of slower modulations of the effective Rabi frequency,
based on optimal control theory, has been proposed in Ref.\cite{exp}.

In summary we propose using optical lattices along with the pulse method
of Garrett et al \cite{Merlin} to achieve remarkable squeezing of the
matter field of atomic vibrations. An important result of our work is the
expression for maximum squeezing (minimum variance) in 
Eqs.(\ref{eq:squeezing_effect}), (\ref{eq:convsqueezing}) and especially
(\ref{eq:absmin}). It appears this strong squeezing can be reached within
existing experimental techniques 
to make the sub-doppler cooling of quantum gases. 
 
An interesting conclusion can be made from our considerations not only
for the optical lattices, but also for the magneto-optical traps.
Namely, the minimum achievable temperature is limited by the size of the
trap. Particularly, for sodium atoms the typical size of the
magneto-optical trap is about $\sim 10^{3}$ of the resonant wavelength and
therefore, the minimum temperature is $\sim 10^{-6}$ of the recoil
temperature. This provides a significant reduction of the temperature.

The work was supported in part by CUNY-PSC Faculty research Award
Program. 
A. Burin acknowledges support in part by Netherlads Organization for
Scientific Research. H. Rabitz acknowledges support from National Science Foundation.  

%DON'T PUT ET AL (HRABITZ): NACHAL'NIKI VSEGDA V KONCE!!!

\end{document}